\documentclass[pra,twocolumn,showpacs]{revtex4-1}
\usepackage{graphicx}
\usepackage{dcolumn}
\usepackage{bm}
\usepackage{bm}
\usepackage{amsmath}
\usepackage{textcomp}
\usepackage[percent]{overpic}
\usepackage{subfig}
\usepackage{amssymb}
\begin{document}

\title{New assessment on the nonlocality of correlation boxes}

\author{A. P. Costa}
\author{Fernando Parisio}
\email{parisio@df.ufpe.br}
\affiliation{Departamento de F\'{\i}sica, Universidade Federal de Pernambuco, 50670-901,Recife, Pernambuco, Brazil}


\begin{abstract}
Correlation boxes are hypothetical systems capable of producing the maximal algebraic violation of Bell inequalities, beyond the quantum bound and without superluminal signaling. The fact that these systems show stronger correlations than those presented by maximally entangled quantum states has been regarded as a demonstration that the former are more nonlocal than the latter. By employing an alternative, consistent measure of nonlocality, we show that this conclusion is not necessarily true. In addition, we find a class of correlation boxes that are less nonlocal than the quantum singlet with respect to the Clauser-Horne-Shimony-Holt inequality, being, at the same time, more nonlocal with respect to the 3322 inequality.
\end{abstract}

\maketitle

\section{Introduction}
To come to grips with quantum nonlocality is a very hard program and
the difficulty is at least twofold. In the first place, nonlocality 
depends upon entanglement, which is by itself difficult to characterize. 
Secondly, and differently from 
entanglement, supplementary information on how nonlocality is to be inferred 
seems to be always necessary. This difference comes from the fact that entanglement can be defined and investigated 
in purely mathematical terms, while nonlocality can be seen as one of the manifestations of entanglement in 
the tangible world. 

The lack of this perception may lead to misleading conclusions, as for example, to claim that a two-qubit state $\rho$ 
is Bell local because it does not violate the Clauser-Horne-Shimony-Holt (CHSH) \cite{chsh} inequality for any setting. It turns out that this
same state may well be nonlocal if the experimental apparatus is changed to investigate the 3322 inequality \cite{33220,3322}. So, it is possible to speak
of the entanglement of $\rho$ alone, but it seems reasonable that an experimental context should be described in order to enable
an assessment of Bell nonlocality. Of course, one can always hope to obtain a finite and exhaustive set of contexts that would account
for a complete description. If we think of the broader viewpoint of generalized measurements, and their infinity of possible Naimark extensions,
this hope may not be easily fulfilled.

An interesting way to investigate to what extent nonlocal behaviors are limited by the structure of quantum mechanics itself, 
is to consider correlation boxes \cite{PR}. These hypothetical physical systems are only constrained by the requirement that they cannot
display correlations enabling superluminal communication. One striking conclusion obtained in \cite{PR} is that the fulfilment of local causality is not sufficient 
to account for the fact that quantum states do not attain the maximal algebraic violation in the CHSH inequality, for, it was shown that nonsignaling boxes
could reach this maximal value. This fact has been acknowledged as a demonstration that correlation boxes could be more nonlocal than maximally 
entangled states. We will provide a critical analysis of the reasoning behind this inference and show that the conclusion may not be so immediate.
It is worth mentioning that although it has been shown that correlation boxes do not have a classical limit \cite{rohrlich} and cannot be used to simulate arbitrary nonsignaling nonlocal correlations \cite{dupuis}, no critical assessment on their degree of nonlocality is available.

Assuming that we intend to evaluate the nonlocal content of a system, given an experimental scenario, the most usual procedure
is to associate larger numeric violations of the corresponding Bell inequality, with a higher degree of nonlocality.
This association has been recently disputed and an alternative quantifier has been proposed along with  
detailed discussions \cite{fonseca,parisio}. Here we limit ourselves to a brief description.
We argue that, although it is true that the larger the violation the more correlated is the system, this inference should not be automatically extended to nonlocality.

In addition to a fixed experimental apparatus, let us consider a fixed set of experimental parameters (e. g. a particular set of angles of 
polarizers). The fact that the numeric violation caused by $\rho$ is larger than that caused by $\sigma$, does not affect in any way
the character of the ``action at a distance" needed to explain the correlations. We, thus, reason that in what concerns nonlocality the two violations 
are equivalent. Given a fixed setup, a state is either local or nonlocal in a Boolean way. By assuming this position, there remains a way to produce a continuous nonlocality hierarchy among physical states: to sum up over all parameters that can be varied within the context of an experiment, attributing weight $1$ to those settings that lead to violation and weight $0$ otherwise.
Accordingly, the state $\rho$ is more nonlocal than $\sigma$ if the former violates local causality, no matter by what extent, for a larger number of experimental configurations than the latter. This leads to the definition of volume of violation within the set ${\cal X }=\{ x_i \}$ of all possible experimental configurations:
\begin{equation}
V(\varrho)=V(\varGamma_{\rho})= \int_{\varGamma_{\rho}}d^n x \;,
\end{equation}
where $\varGamma_{\rho}$ is the subset of ${\cal X }$ containing all violating settings and $d^n x=\mu(x_1,\dots, x_n)dx_1\dots dx_n$.
The measure $\mu$ is such that every configuration is equally important.
In many cases, the relative quantity $v=V(\rho)/V_T$, with $V_T=\int_{\cal X}d^n x$, is more relevant, since it can be interpreted as a probability
of violation when a random, unbiased choice of settings is made. This very probability has been defined in the context of entanglement detection via random
local measurements \cite{random}.

The nonlocal content of a state as given by the volume of violation has been applied to a problem that became 
known as the ``anomaly'' in the nonlocality of two three-level systems \cite{anomaly}. It consists in the fact that the Collins-Gisin-Linden-Massar-Popescu (CGLMP) inequality \cite{cglmp} is maximally violated by a state that is not maximally entangled for two entangled qutrits. It turns out that the supposed anomaly disappears when the volume of violation is employed, that is, $V$ attains its maximum for the maximally entangled state \cite{fonseca}. 

In this work we use the volume of violation to assess the nonlocality of a parametrized family of correlation boxes which always produce maximal 
algebraic violations in the CHSH inequality for a particular experimental setting.
As usual, we will consider the correlation function $E$ associated with measurements of arbitrary components of two spin-1/2 .
If the spin component is measured in direction ${\bf a}$ for the first subsystem and in direction ${\bf b}$ for the second subsystem,
$E$ is given by
\begin{equation}
E({\bf a},{\bf b})=P_{{\bf a}{\bf b}}(+,+)+P_{{\bf a}{\bf b}}(-,-)-P_{{\bf a}{\bf b}}(+,-)-P_{{\bf a}{\bf b}}(-,+)\;,
\end{equation}
where $P_{{\bf a}{\bf b}}(\cdot, \cdot)$ is the probability of a particular outcome. For the spherically symmetric 
two-qubit state $|\Psi_S\rangle=(|+-\rangle-|-+\rangle)/\sqrt{2}$ we get
\begin{equation}
E({\bf a},{\bf b})=E_S(\theta)=-{\bf a}\cdot{\bf b}=-\cos \theta_{ab}\;.
\end{equation}
\section{Popescu-Rohrlich Box}
\label{PRsec}
Back in 1994 Popescu and Rohrlich reversed one of the most persisting questions in modern physics \cite{PR}:  
``Rather than ask why quantum correlations violate the CHSH inequality, 
we might ask why they do not violate it {\it more}''. A way to understand the elementary principles behind the
Tsirelson bound, only reached by maximally entangled two-qubit states, is to look for hypothetical systems
that go beyond this limit, yet, respecting the nonsignaling requirement. The fact that this is not a mathematical 
impossibility is compelling by itself. 

\begin{figure}[t]
\includegraphics[width=4cm]{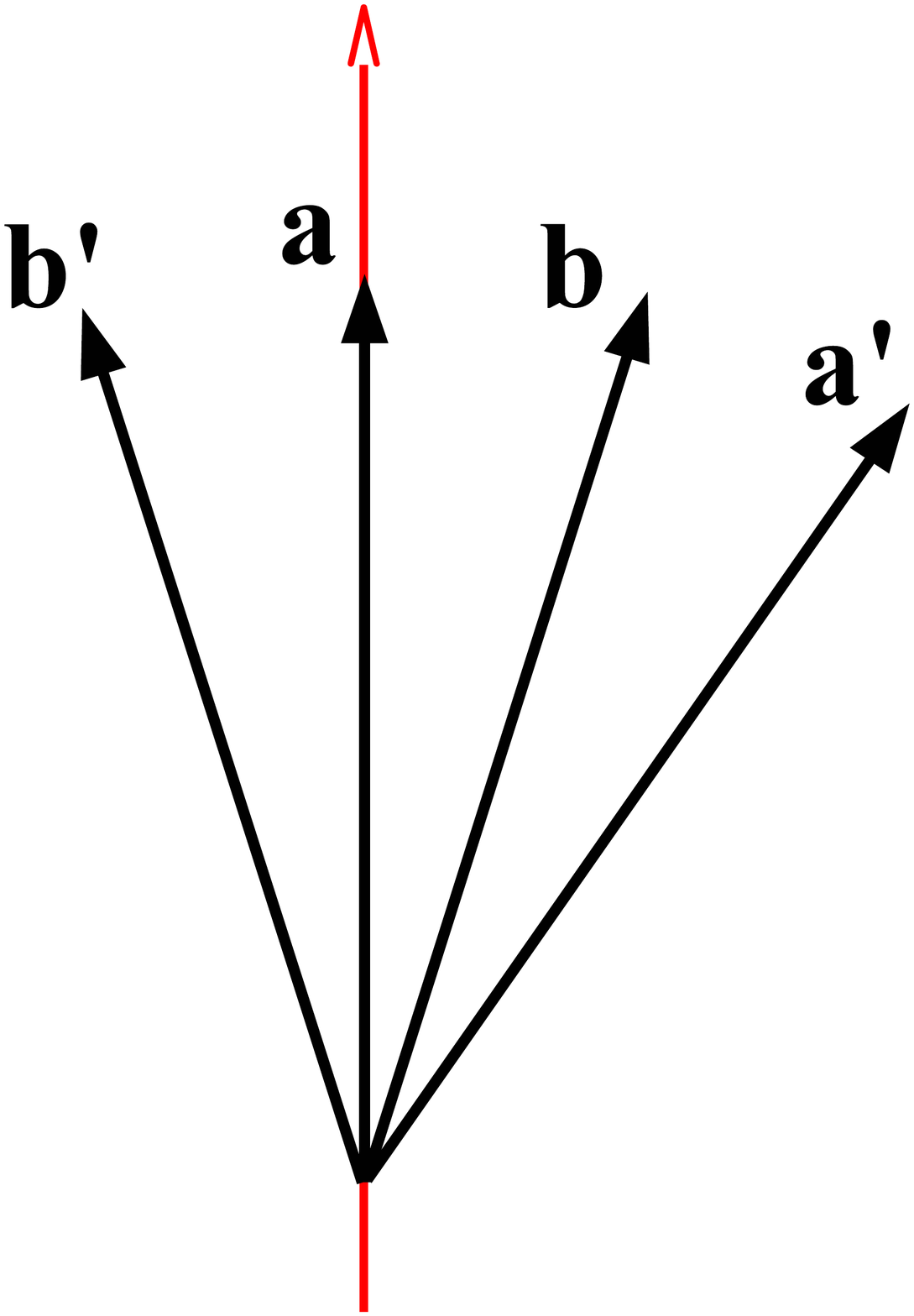}
\caption{(color online) In-plane configuration with $\theta_{ab}=\theta_{ab'}=\theta_{a'b}=\pi/12$ and $\theta_{a'b'}=\pi/4$.
This gives $E_{PR}(\theta_{ab})+E_{PR}(\theta_{ab'})+E_{PR}(\theta_{a'b})-E_{PR}(\theta_{a'b'})=4$, the larger possible
algebraic value and above the maximum value attained by the quantum singlet ($2\sqrt{2}$).  }
\label{fig1}
\end{figure}

It is worth mentioning that, some times, correlation boxes are seen too schematically,
as something deprived from any geometrical feature. We emphasize that we treat the boxes as physical, though gedanken systems. 
Even for hypothetical two level structures, for instance, measurements in any direction should be conceivable, as 
is the case of actual spins. To keep a close analogy 
with the singlet (a maximally entangled state) and to simplify the discussion, the authors of \cite{PR} assumed that their
supraquantum \cite{comment} system was spherically symmetric. The correlation function they proposed reads
\begin{displaymath}\label{funcaopr}
E_{PR}(\theta) = \left\{ \begin{array}{ll}
1  &\;\;\;\mbox{for}\;\;\; 0 \leq \theta <\pi/6\\
\\
-\dfrac{24}{\pi}\theta + 5 &\;\;\;\mbox{for}\;\;\; \pi/6 \leq \theta < \pi/4\\
\\
-1 & \;\;\;\mbox{for}\;\;\; \pi/4 \leq \theta < \pi/3\\
\\
\dfrac{6}{\pi}\theta - 3 &\;\;\;\mbox{for}\;\;\; \pi/3 \leq \theta < 2\pi/3\\
\\
1 &\;\;\;\mbox{for}\;\;\; 2\pi/3 \leq \theta < 3\pi/4\\
\\
-\dfrac{24}{\pi}\theta + 19 &\;\;\;\mbox{for}\;\;\; 3\pi/4 \leq \theta < 5\pi/6\\
\\
-1 &\;\;\;\mbox{for}\;\;\; 5\pi/6 \leq \theta \leq \pi
\end{array} \right .
\end{displaymath}
where $\theta$ is the angle between the two measurement directions.
Although contrived, as the authors acknowledge, it serves the purpose of showing that it is, in principle, possible to attain
the maximum algebraic violation for the CHSH inequality
\begin{equation}
\label{chsh}
-2\le E(\theta_{ab})+E(\theta_{ab'})+E(\theta_{a'b})-E(\theta_{a'b'})\le 2\;,
\end{equation}
without signaling. To see this we only need to consider the in-plane configuration shown in fig. \ref{fig1}.
\begin{figure*}
\includegraphics[width=5.2cm,angle=0]{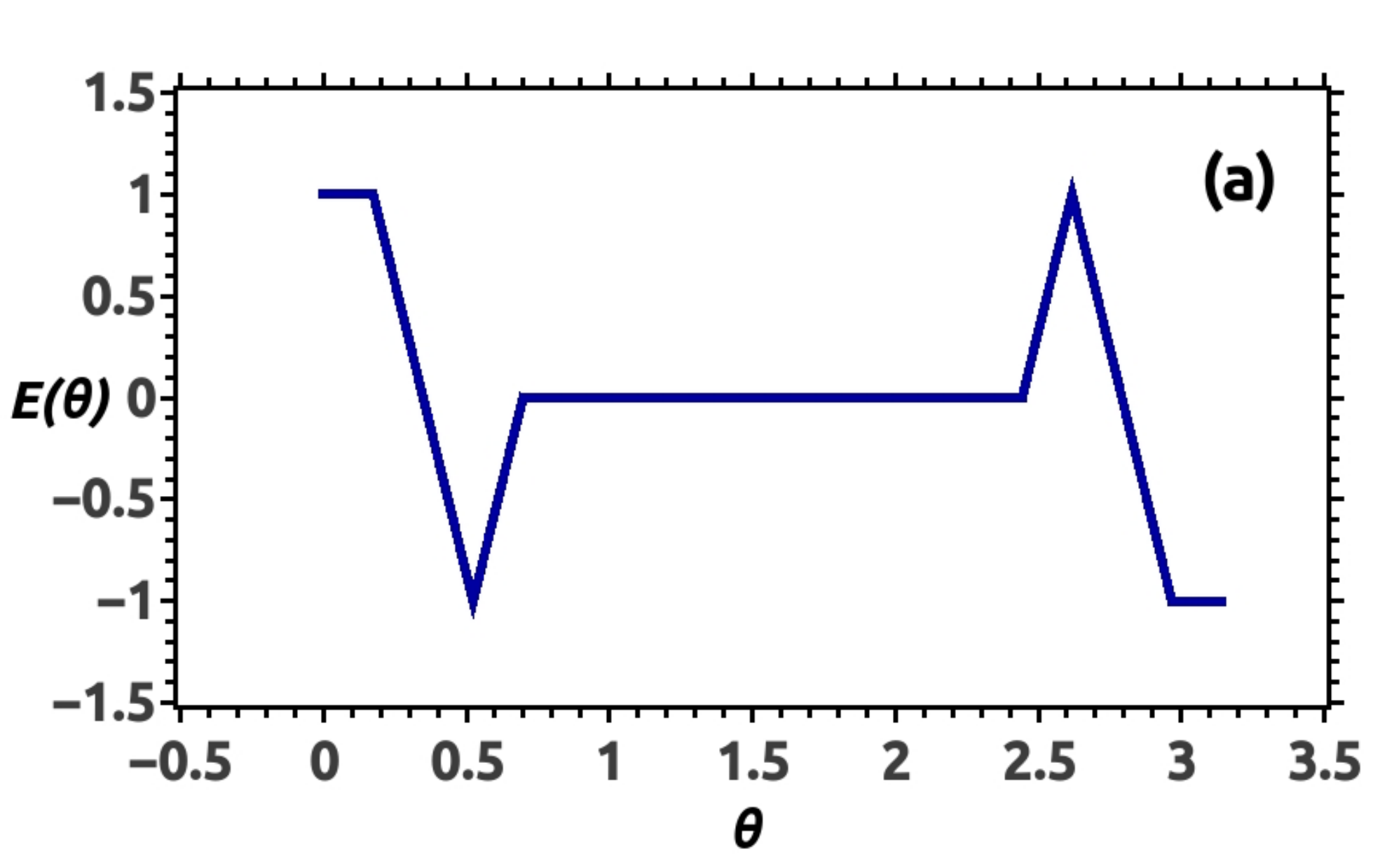}
\includegraphics[width=5.2cm,angle=0]{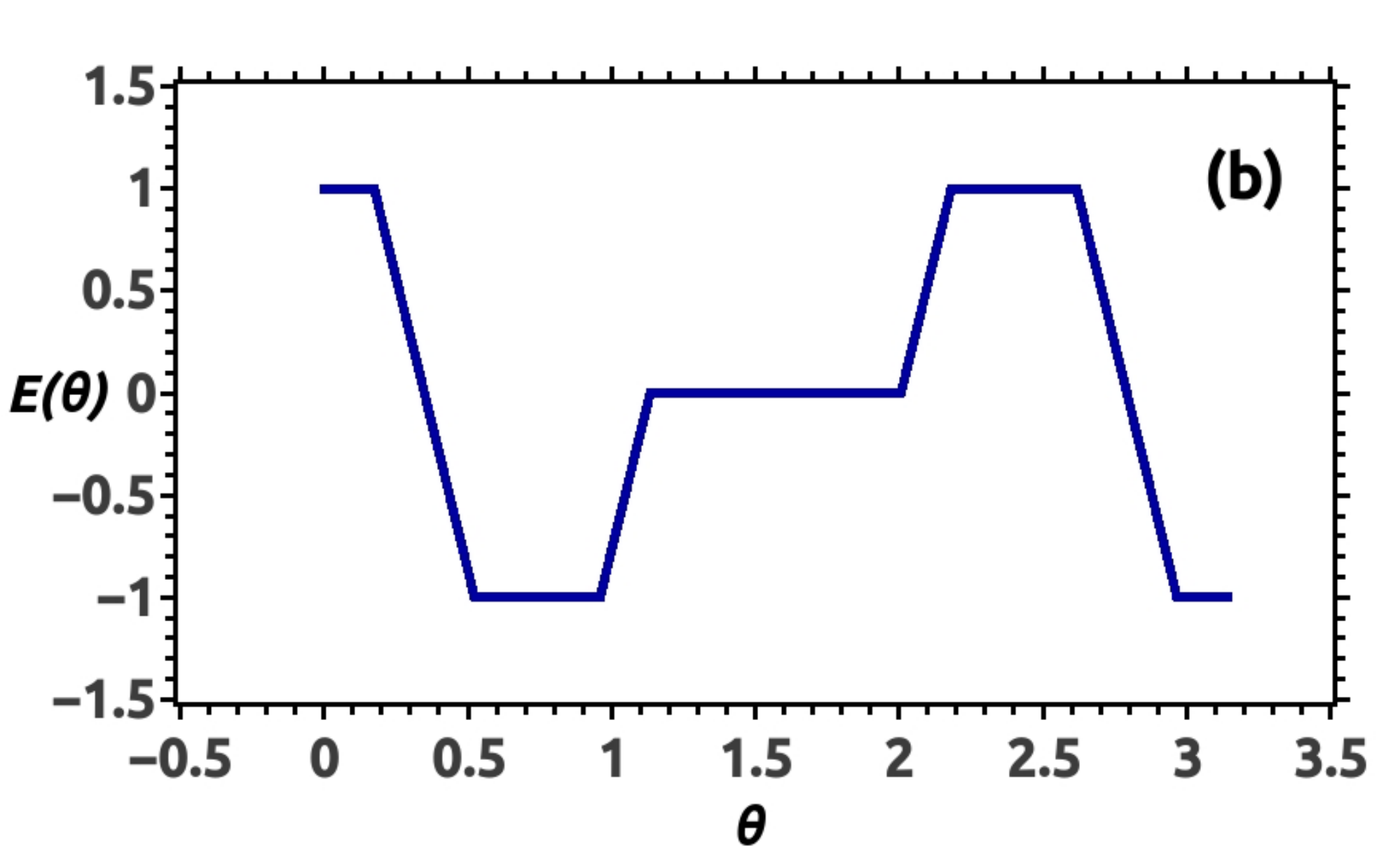}
\includegraphics[width=5.2cm,angle=0]{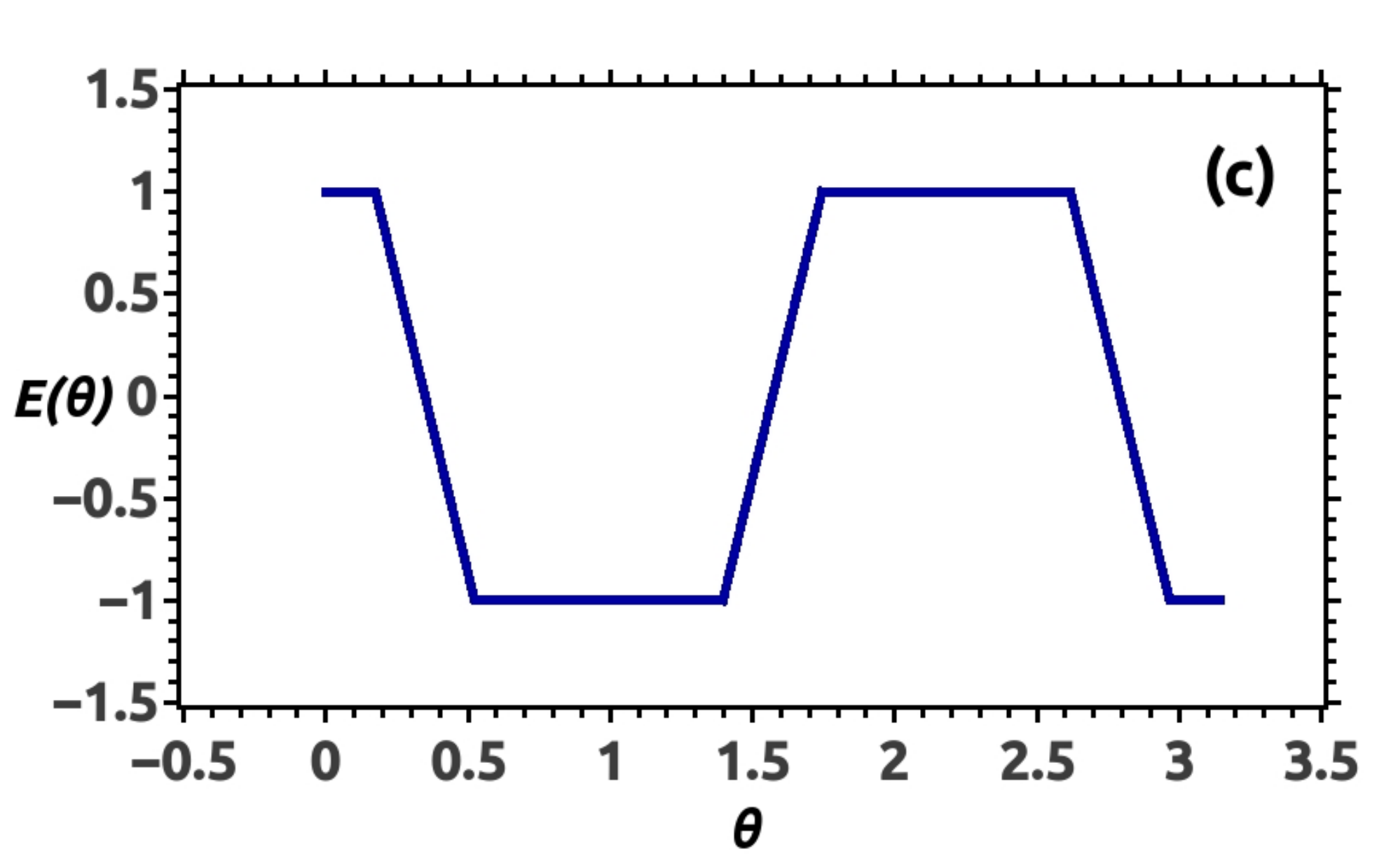}
\caption{(color online) ``Supraquantum'' correlation functions $E_{\lambda}(\theta)$ for (a) $\lambda=\pi/6$, (b) $\lambda=11\pi/36$, and  (c) $\lambda=4\pi/9$.}
\label{fig2}
\end{figure*}

It is simple to numerically compute the volume of violation of the quantum singlet and PR-box, for a test of the CHSH inequality.
Firstly, due to the spherical symmetry of both systems one can choose one out of the four involved directions to be fixed, without any
loss of generality. Let this direction be ${\bf a}={\bf z}$.
Thus, in the most general measurement situation we have:
\begin{eqnarray*}
{\bf a} & = & (0, 0, 1), \\
{\bf b} & = & (\sin\theta_b \cos\phi_b, \sin\theta_b \sin\phi_b, \cos\theta_b), \\
{\bf a'} & = & (\sin\theta_{a'} \cos\phi_{a'}, \sin\theta_{a'} \sin\phi_{a'}, \cos\theta_{a'}), \\
{\bf b'} & = & (\sin\theta_{b'} \cos\phi_{b'}, \sin\theta_{b'} \sin\phi_{b'}, \cos\theta_{b'}).
\end{eqnarray*}

This determines the angles that directly enter in the inequality, e. g., $\theta_{ab}=\arccos({\bf a}\cdot{\bf b})=\theta_b$,
\begin{eqnarray}
\nonumber
\theta_{a'b}=\arccos({\bf a'}\cdot{\bf b})=\\
\nonumber
\arccos[\sin \theta_{a'}\sin \theta_b\cos(\phi_{a'}-\phi_b)+\cos \theta_{a'}\cos \theta_b]\;, 
\end{eqnarray}
and so on. Next one draws random values for $(\phi_b,\theta_b)$,  $(\phi_{a'},\theta_{a'})$,
and $(\phi_{b'},\theta_{b'})$, uniformly distributed over each sphere. These values feed $\theta_{AB}$ with $A=a,a'$ and $B=b,b'$ which, in turn,
are used in a Monte Carlo integration over a 6-dimensional bounded manifold with total volume $V_T=(4\pi)^3$.

According to our calculations the ratio between the volume of the singlet and the total volume is $V_S/V_T=v_S\approx 0.070799$, which is in excellent agreement with
the closed analytical result $(\pi-3)/2\approx 7.0796\%$ obtained in the supplemental material of \cite{random}. Here we will not be concerned
with relabelling of axes \cite{random} because this would give the same multiplicative factor for all volumes, being, therefore, irrelevant for comparisons involving the same inequality.
For the PR-box we obtained $V_{PR}/V_T=v_{PR} \approx 0.180717$, which indicates that
the Popescu-Rohrlich Box is, indeed, more nonlocal than the quantum singlet in the CHSH scenario. The question arises whether or not
this is a logical necessity in general. Does a spherically symmetric correlation box giving the maximal algebraic violation 
for a certain Bell inequality, always present a volume of violation larger than that of the quantum singlet? 
As we will see in the next section, the answer is negative.
\section{Nonlocality of a family of correlation boxes}
In this section we provide a family of correlation boxes parametrized by a real number $\lambda \in [\pi/6,4\pi/9]$, that is, $30^{\rm o}\le \lambda \le 80^{\rm o} $.
For {\it all} values of $\lambda$ in this interval the boxes are spherically symmetric ($\Rightarrow$ nosignaling) and produce the maximal
algebraic violation of inequality (\ref{chsh}). Therefore, according to the criterion that associates a higher degree of
nonlocality with larger numeric violations, these boxes are more nonlocal than the quantum singlet, and, as nonlocal as
the original PR-box for all $\lambda$'s. The supraquantum correlation function $E_{\lambda}(\theta)$ for these systems, let us call them $\lambda$-boxes, 
is defined by 
\begin{displaymath}\label{funcaoparametrizada}
E_{\lambda}(\theta) = \left\{ \begin{array}{ll}
1 & \;\;\;\mbox{for}\;\;\; 0 \leq \theta < \pi/18\\
\\
-\dfrac{18}{\pi}\theta + 2 &\;\;\;\mbox{for}\;\;\; \pi/18 \leq \theta < \pi/6\\
\\
-1 &\;\;\;\mbox{for}\;\;\; \pi/6 \leq \theta <\lambda\\
\\
\dfrac{18}{\pi}(\theta - \lambda) - 1 &\;\;\;\mbox{for}\;\;\; \lambda \leq \theta < \lambda + \pi/18\\
\\
0 &\;\;\mbox{for}\;\; \lambda + \pi/18 \leq \theta < 17\pi/18 - \lambda\\
\\
\dfrac{18}{\pi}(\theta + \lambda) - 17 &\;\;\mbox{for}\;\; 17\pi/18 - \lambda \leq \theta < \pi - \lambda\\
\\
1 &\;\;\;\mbox{for}\;\;\; \pi - \lambda \leq \theta < 5\pi/6\\
\\
-\dfrac{18}{\pi}\theta + 16 &\;\;\;\mbox{for}\;\;\; 5\pi/6 \leq \theta < 17\pi/18\\
\\
-1 &\;\;\;\mbox{for}\;\;\; 17\pi/18 \leq \theta \leq \pi
\end{array} \right .
\end{displaymath}
In fig. \ref{fig2} we show $E_{\lambda}(\theta)$ as a function of $\theta$ for (a) $\lambda=\pi/6 \;(30^{\rm o})$, (b) $\lambda=11\pi/36 \;(55^{\rm o})$, and (c) $\lambda=4\pi/9 \;(80^{\rm o})$.
It is crucial to note that, no matter the value assumed by $\lambda$, for  $\theta_{ab}=\theta_{ab'}=\theta_{a'b}=\pi/18$ and $\theta_{a'b'}=\pi/6$, the Bell function in inequality (\ref{chsh}) reaches the value 4. 

\subsection{Clauser-Horne-Shimony-Holt Nonlocality}
\begin{figure}[t]
\includegraphics[width=7cm,angle=0]{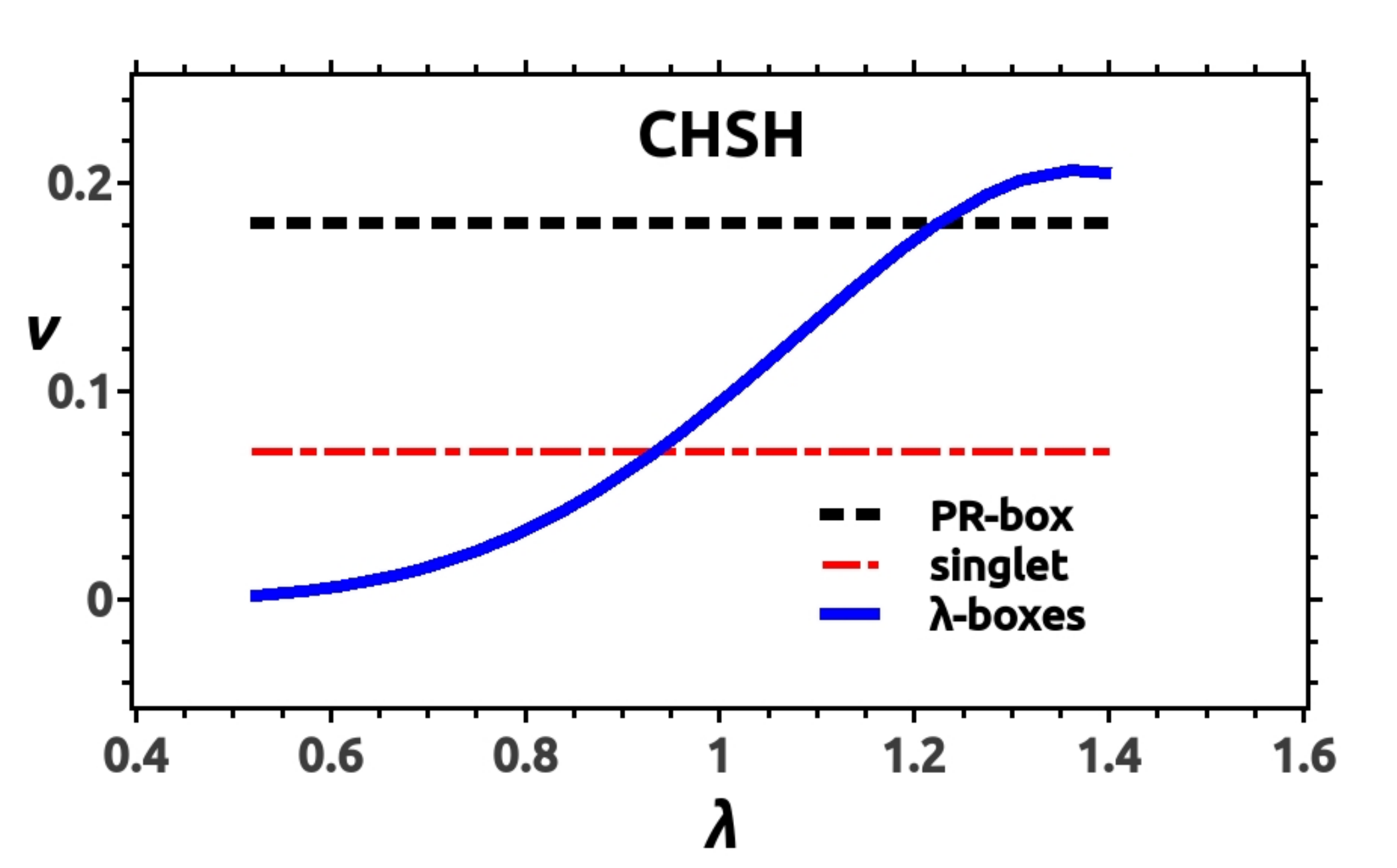}
\caption{(color online) Nonlocality, as given by the ratio $v=V/V_T$, of the quantum singlet (dot-dashed line), PR-box (dashed line), and $\lambda$-boxes (continuous curve) for the CHSH inequality. The latter are less nonlocal than the singlet if $\lambda<\tilde{\lambda}_{CHSH}=0.934$ and more nonlocal than the PR-box if $\lambda>1.225$. }
\label{fig3}
\end{figure}
We employed the same procedure as in Sec. \ref{PRsec} to determine the volume of violation, as a function of $\lambda$, of these boxes. The results are depicted in fig. \ref{fig3}, from $\lambda=\pi/6\approx 0.524$ to $\lambda=4\pi/9\approx 1.396$. The two horizontal lines give the volumes of the quantum singlet (dot-dashed) and PR-box (dashed), while the continuous curve describes the amount of nonlocality associated with the $\lambda$-family of correlation boxes. As it is clear, there is a value of $\lambda$ ($\lambda=\tilde{\lambda}_{CHSH}\approx 0.934$) below which, the $\lambda$-boxes are less non-local than the singlet, according to our criterion. Recall that, for all $\lambda$'s the maximal value of the CHSH function is 4. In addition, they are more nonlocal than the original PR-box for $\lambda>1.225$, showing that a higher degree of nonlocality may have little to do with maximal violations for specific configurations.

\subsection{3322 Nonlocality}
\begin{figure}[t]
\includegraphics[width=7cm,angle=0]{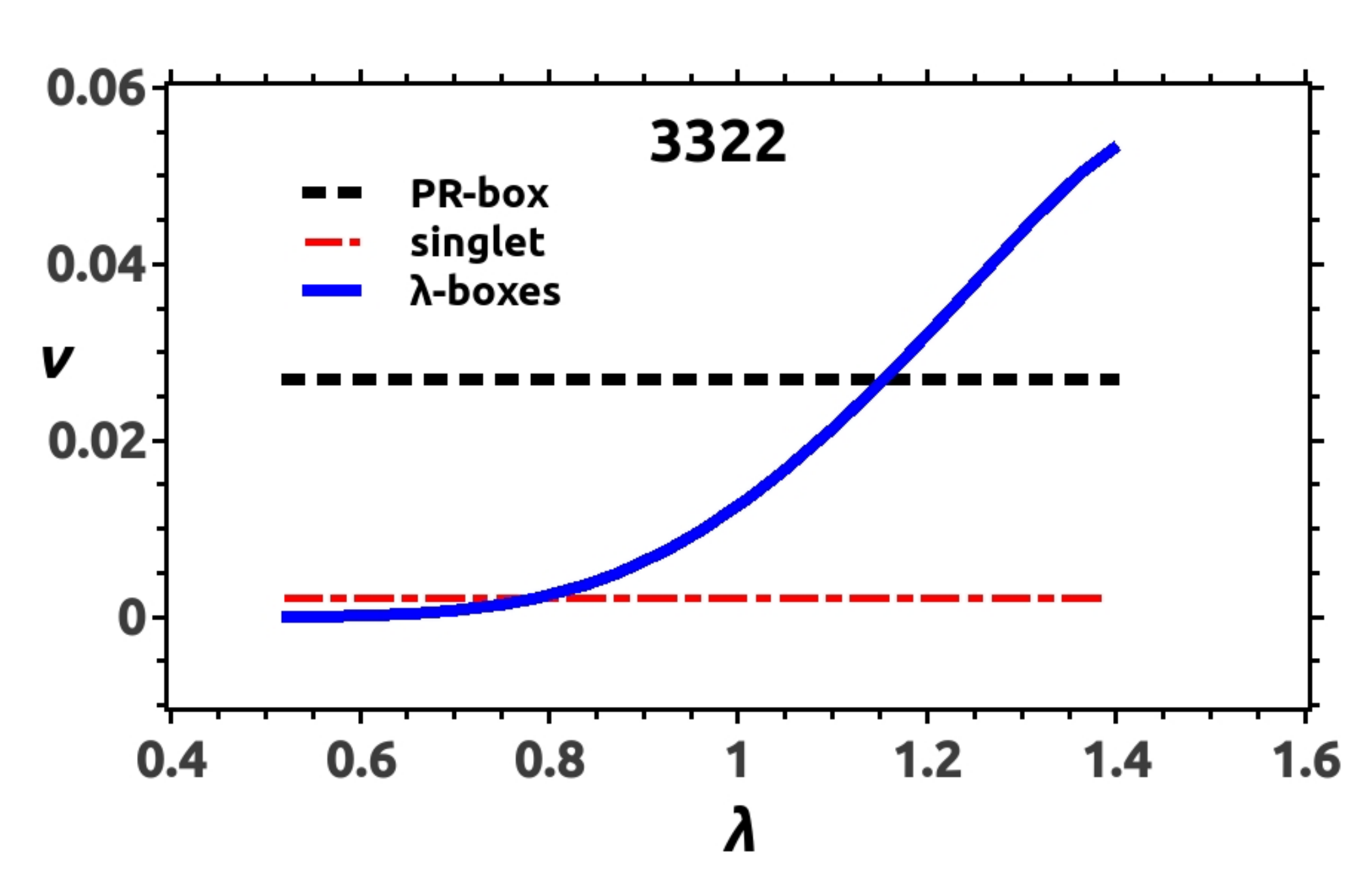}
\caption{(color online) Nonlocality, as given by the ratio $v=V/V_T$, of the quantum singlet (dot-dashed line), PR-box (dashed line), and $\lambda$-boxes (continuous curve) for the 3322 inequality. The latter are less nonlocal than the singlet if $\lambda<\tilde{\lambda}_{3322}=0.788$ and more nonlocal than the PR-box if $\lambda>1.154$.}
\label{fig4}
\end{figure}

We proceed to make an analogous numeric investigation within a relevant, distinct context, for which three observables per site may be selected. Let us consider
that these observables can be fully characterized by three, otherwise arbitrary, unit vectors per party, $({\bf a}, {\bf a'},{\bf a''})$ and $({\bf b}, {\bf b'},{\bf b''})$.
The tight Bell inequality involving two qubits for this physical situation has been derived in \cite{33220,3322} and reads
\begin{eqnarray}
\label{3322}
\nonumber
I_{3322}  =  -E({\bf a})-E({\bf a'})+E({\bf b})+E({\bf b'})+E({\bf a},{\bf b})+\\
\nonumber
E({\bf a},{\bf b'})+E({\bf a},{\bf b''}) +E({\bf a'},{\bf b})+E({\bf a'},{\bf b'})-\\
\nonumber
 E({\bf a'},{\bf b''})+E({\bf a''},{\bf b})-E({\bf a''},{\bf b'}) \leq 4,
\end{eqnarray}
where the correlations with a single argument refer to measurements on one of the parties only, and vanish for spherically symmetric states.
The Monte Carlo integrations are now over a 10-dimensional bounded manifold with a total volume of $V_T=(4\pi)^{5}$, with spherical symmetry already taken into account.
In the present case, the maximal violation by a quantum state corresponds to the value 5, and, although the $\lambda$-boxes were designed to produce the maximal
algebraic violation in the CHSH context, they also go far beyond quantum mechanics here, yielding $I_{3322}^{max}\approx8.0$. Our results are plotted in fig. \ref{fig4}.
Here $v_S=2.17 \times 10^{-3}$ and $v_{PR}=2.69 \times 10^{-2}$, again, without considering relabelings of observables.
They are qualitatively similar to those of fig. \ref{fig3}, with the boxes passing from less nonlocal to more nonlocal than the singlet as $\lambda$ grows.  There are, however, important differences: (i) the singlet becomes less nonlocal earlier ($\lambda=\tilde{\lambda}_{3322} \approx 0.788$) and (ii) the nonlocality for $\lambda=4\pi/9$ is about twice as that of the PR-box (see the rightmost part of fig. \ref{fig4}).

The fact that $\tilde{\lambda}_{CHSH}\ne\tilde{\lambda}_{3322}$ leads to a quite interesting situation which illustrates the difficulties involved in making statements on the nonlocality of a system, without mentioning the context. For $\tilde{\lambda}_{3322}< \lambda<\tilde{\lambda}_{CHSH}$ we have the quantum singlet being more nonlocal than the $\lambda$-boxes with respect to the CHSH inequality, and, at the same time, less nonlocal than the {\it same} $\lambda$-boxes for 3322 tests. Although the CHSH inequality is older and better known, it can not be claimed that it is more relevant than the 3322 inequality in any obvious sense. It is crucial to note that these two inequivalent inequalities correspond to all non-trivial facets of the local polytope \cite{3322}. This means that these two tight inequalities are exhaustive and are at the same footing, at least as far as we are restricted to von Neumann measurements. 
An analogous phenomenon arises in the rightmost part of panels \ref{fig3} and \ref{fig4} involving the PR-box and the $\lambda$-boxes for $1.154< \lambda<1.225$.

Note that this apparent paradox is not due to our particular way to quantify nonlocality. It is possible to devise a different family of boxes which always go beyond the Tsirelson bound, without, however, reaching the algebraic maximum. If we directly use the traditional criterion, the same conflict may appear when the 3322 context is considered.

\section{Conclusion}
We employed the concept of volume of violation to quantify the nonlocality of states submitted to CHSH and 3322 measurements. In both scenarios we found that the Popescu-Rohrlich box is more nonlocal than the quantum singlet. However, this is shown to be incidental. We defined a continuous family of correlation boxes, keeping the essential features of the Popescu-Rohrlich supraquantum system, namely,  spherical symmetry, nonsignaling, and maximal algebraic violation of the CHSH inequality. We  showed that the fact that a system presents a larger maximal violation than that of the singlet for a particular set of experimental parameters, does not necessarily imply that it is more nonlocal, according to our proposed criterion.

The high degree of nonlocality presented by the $\lambda$-boxes in the case of three observables per site rises a question on maximal violations and spherical symmetry. Is there a nonsignaling box which attain the algebraic maximum of 12 in the 3322 inequality? If the answer is positive then the referred box must not be spherically symmetric, since the four selfcorrelations appearing in the 3322 inequality would vanish, restricting the maximum to 8, which was indeed reached by the $\lambda$-boxes. If, on the contrary, there are boxes which produce even larger violations, then, in general, spherical symmetry, besides not being a necessary, is also not a sufficient condition for maximal nonlocality of pure states.

More interestingly, the $\lambda$-boxes and the quantum singlet may have a different nonlocality hierarchy depending on which tight Bell inequality one is investigating, for $\lambda$ between $45.1^{\rm o}$ and $53.4^{\rm o}$. So, can we make sense of questions like: which is more nonlocal, the singlet or the $\lambda$-boxes, for $\tilde{\lambda}_{3322}< \lambda<\tilde{\lambda}_{CHSH}$? It seems that there is no way to pose this kind of question without referring to a particular experimental context. There may be one way out: to calculate the volume of violation directly in the space of probabilities $\{p(ab|xy)\}$ (the volume of nonlocal behaviors), without referring to a particular Bell inequality. Whether or not this is indeed sufficient is not an easy question. What about hidden nonlocality, which requires intermediate, but local measurements to be revealed \cite{popescu}? How to deal with positive operator valued measurements and the possible nonlocality concealed in each Naimark extension \cite{vertesi, cabello}? These questions seem to be worth of some thought.

\begin{acknowledgements}
The authors thank E. A. Fonseca for a critical reading of this manuscript. Financial support from Conselho Nacional de Desenvolvimento Cient\'{\i}fico e Tecnol\'ogico (CNPq), Coordena\c{c}\~ao de Aperfei\c{c}oamento de Pessoal de N\'{\i}vel Superior (CAPES), and Funda\c{c}\~ao de Amparo \`a Ci\^encia e Tecnologia do Estado de Pernambuco (FACEPE) is acknowledged.
\end{acknowledgements}


\begin{thebibliography}{90}
\bibitem{comment} We prefered the term ``supraquantum'' instead of ``superquantum'', 
employed by Popescu and Rohrlich, because the former indicates something that transcends 
quantum mechanics, while the latter may be misinterpreted as ``strong but still quantum''.
\end{thebibliography}
\end{document}